\documentclass[twocolumn,showpacs,preprintnumbers,amsmath,amssymb,superscriptaddress]{revtex4}

\usepackage{graphicx}
\usepackage{dcolumn}
\usepackage{bm}
\usepackage{units}

\begin{document}


\title{Strongly birefringent cut-wire pair structure as negative index wave plates at THz frequencies}

\author{P. Weis}
\email{weis@rhrk.uni-kl.de}
\affiliation{Department of Physics and Research Center OPTIMAS, University of Kaiserslautern, Germany
}%
\affiliation{Fraunhofer Institute for Physical Measurement Techniques IPM, Freiburg, Germany
}%
\author{O. Paul}
\affiliation{%
Department of Physics and Research Center OPTIMAS, University of Kaiserslautern, Germany
}%
\author{C. Imhof}
\affiliation{%
Department of Electrical and Computer Engineering, University of Kaiserslautern, Germany
}%
\author{R. Beigang}
\affiliation{Department of Physics and Research Center OPTIMAS, University of Kaiserslautern, Germany
}%
\affiliation{Fraunhofer Institute for Physical Measurement Techniques IPM, Freiburg, Germany
}%
\author{M. Rahm}
\affiliation{Department of Physics and Research Center OPTIMAS, University of Kaiserslautern, Germany
}%
\affiliation{Fraunhofer Institute for Physical Measurement Techniques IPM, Freiburg, Germany
}%

\date{\today}

\begin{abstract}
We report a new approach for the design and fabrication of thin wave plates with high transmission in the terahertz (THz) regime. The wave plates are based on strongly birefringent cut-wire pair metamaterials that exhibit refractive indices of opposite signs for two orthogonal polarization components of an incident wave. As specific examples, we fabricated and investigated a quarter- and a half-wave plate that revealed a peak intensity transmittance of 74\% and 58\% at 1.34~THz and 1.3~THz, respectively. Furthermore, the half wave plate displayed a maximum figure of merit (FOM) of 23 at 1.3 THz where the refractive index was $-1.7$. This corresponds to one of the highest FOMs reported at THz frequencies so far. The presented results evidence that negative index materials enter an application stage in terms of optical components for the THz technology.
\end{abstract}

\maketitle

In the last decade, metamaterials have developed to one of the hot topics in the research field of optics and electromagnetics. Although a great deal of scientific interest initially originated from the pursuit of materials with negative refraction \cite{veselago:1968,pendry:2000}, it became evident that this new class of artificial materials offers potent tools for tailoring the electromagnetic properties of light in general \cite{Pendry:2006,Schurig:2006}. Since metamaterials can be designed to provide both electric and magnetic response to external electromagnetic fields, it is possible to fabricate artificial functional structures with distinctive optical properties. Though in principle, metamaterials can be conceived to operate in any frequency range, they promise to be highly beneficial as optical components in the terahertz (THz) range where conventional optics is not readily available. In fact, most natural materials lack of electric or magnetic response to THz radiation. In this context, metamaterials paved the way to optical \cite{chen:2007} and electric modulators \cite{chen:2006,paul:2009},  phase shifters \cite{chen:2009}, bandpass filters \cite{paul:2009b} and perfect absorbers \cite{landy:2008,Tao:2008,landy:2009} in the THz range.

Recently, Strikwerda et al.\ demonstrated that birefringent electric split-ring resonators and meanderline structures can serve as efficient quarter wave plates in the THz frequency range \cite{strikwerda:2009}. In this Letter, we propose and experimentally verify an alternative approach for the construction of both THz quarter wave plates (QWPs) and half-wave plates (HWPs) based on negative index metamaterials. In contrast to the perfect lens \cite{pendry:2000}, where highly isotropic negative index materials are required, we exploited the strong birefringence of cut-wire pairs \cite{imhof:2007,zhou:2006,gundogdu:2008,dolling:2005} to realize a medium that changes the sign of the refractive index for two orthogonal polarization states. The QWP and the HWP displayed a peak intensity transmittance of 74\% and 58\%, respectively. For the negatively refracted polarization state a maximum figure of merit (FOM) of 23 could be observed at 1.3 THz ($n'=-1.7$) which corresponds to one of the highest FOMs reported at THz frequencies.

An efficient way to create wave plates is the use of birefringent materials. In such media, the refractive index is dependent on the propagation direction (wave vector) and the angle of the polarization vector of the incident wave with respect to the optical axis of the material. This means that, at a given propagation direction, an incoming wave with polarization components parallel and orthogonal to the optical axis experiences different phase advances for both polarization components. The phase delay between the orthogonal polarization components is given by
\begin{equation}
\Delta \phi = \frac{2\pi d \nu}{c} \cdot (n'_\bot - n'_{||})
\end{equation}
where $\nu$ denotes the wave frequency, $c$ the vacuum light velocity and $n'_\bot$ and $n'_{||}$ the refractive indices for the orthogonal and the parallel polarization component, respectively.
At a phase difference of $90^\circ$ (QWP) the polarization state of light changes from linear to elliptical or circular polarization and vice versa.  On the other hand, the orientation of the wave vector of linearly polarized light can be rotated by twice the angle between the polarization vector and the optical axis when the phase difference is $180{^\circ}$ (HWP).

While conventional birefringent media exist for the optical frequency range, such components are not naturally availabe for the THz range and must be artificially assembled. In this regard, metamaterials provide suitable means for the implementation of strongly birefringent components that can serve as wave plates with arbitrary phase delay. For this purpose, we employed a cut-wire pair design. It is well known, that a parallel plate configuration can support a negative index of refraction for waves polarized parallel to the cut wires and a positive refractive index for orthogonally polarized waves and hence is strongly birefringent \cite{imhof:2007,zhou:2006b}.
\begin{figure}[b]
\centering
   \includegraphics[width=\columnwidth]{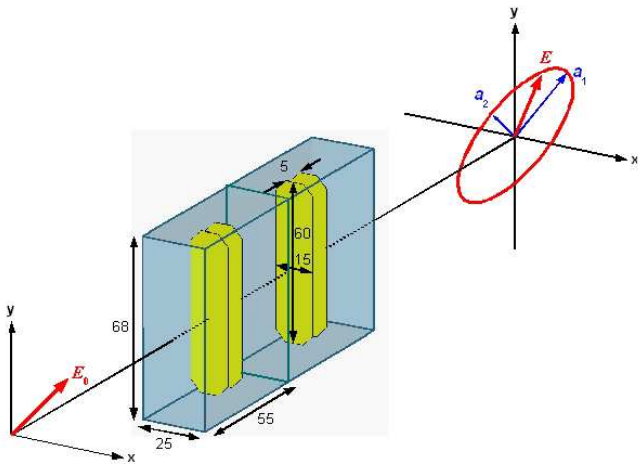}
   \caption{Schematic of a two-unit-cell cut wire pair structure with indicated dimensions in micrometer and polarization ellipse}
    \label{fig:setup}
\end{figure}

As a result of the numerical calculations it became obvious that a one-unit-cell cut-wire pair structure was an appropriate design for a QWP whereas a HWP could be readily obtained by adding a second cut-wire pair unit cell. The dimensions of the fabricated structures are shown in Fig.\ \ref{fig:setup}. We fabricated the metamaterials by a multilayer process composed of layers of bencocyclobutene (BCB) 3022-63 and copper. The fabrication method was similar to the one reported in Ref.\ \cite{paul:2008} and we refer the reader for details. As a result, we obtained flexible, free standing metamaterial membranes with one unit cell for the QWP and two unit cells for the HWP.
\begin{figure}[]
   \begin{center}
   \includegraphics[width=\columnwidth]{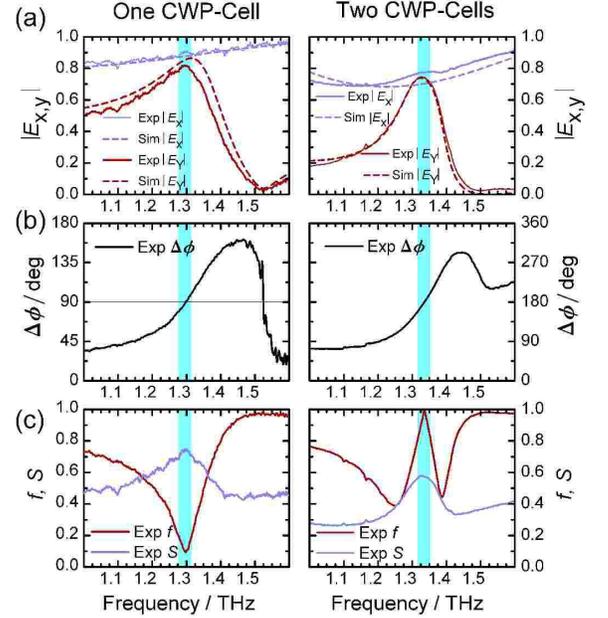}
   \end{center}
   \caption{ (a) Comparison of the simulated and measured amplitude transmittance $|E_x|$ and $|E_y|$ of a QWP (left column) and a HWP (right column), (b) phase difference $\Delta \phi$ and (c) flattening $f$ and intensity transmittance $S$. The metamaterial was composed of cut-wire pair (CWP) unit cells. The shadowed regions indicate the spectral bandwidth of the wave plates as defined in the text.}
    \label{fig:results}
\end{figure}

The proposed wave plates have been characterized by analyzing the polarization state of the
transmitted wave when a linearly polarized wave was incident normally to the plates at a
polarization angle of $45^\circ$ with respect to the x-axis (see Fig.\ \ref{fig:setup}). The polarization state can be quantified by the flattening $f$ of the polarization ellipse which is
defined via the two principle axes $\textbf{\emph{a}}_1$ and $\textbf{\emph{a}}_2$ by
\begin{align*}
f = 1- \left(\frac{|\textbf{\emph{a}}_1|}{|\textbf{\emph{a}}_2|}\right)^{\pm1}, \quad 0\le f \le 1
\end{align*}
where the plus sign is used if $|\textbf{\emph{a}}_1|<|\textbf{\emph{a}}_2|$ and the minus sign
otherwise.
The flattening is $f=0$ for circular polarized waves and $f=1$ for linearly polarized waves. The
principle axes of the polarization ellipse are related to the complex amplitude vector
$\textbf{\emph{E}}$ of the wave by $ \textbf{\emph{E}} =
(\textbf{\emph{a}}_1+i\textbf{\emph{a}}_2)e^{\frac{i}{2}\gamma} $ with $\gamma =
\text{arg}(\textbf{\emph{E}}^2)$. Reversely, the principal axes and thus the flattening can be
calculated from the measured amplitude vector $\textbf{\emph{E}}$ by $\textbf{\emph{a}}_1  =
\text{Re}[\textbf{\emph{E}}e^{-\frac{i}{2}\gamma}]$ and $\textbf{\emph{a}}_2  =
\text{Im}[\textbf{\emph{E}}e^{-\frac{i}{2}\gamma}]$, respectively.

In the experiments we used a standard THz time domain system which operates with linearly polarized
waves. Due to the in-plane mirror symmetry of the cut wire structure, there is no
cross-polarization in the transmission response and the x-and y-component of the transmitted field
vector $\textbf{\emph{E}}=(E_x,E_y)$ can be measured independently by simply rotating the sample.
Furthermore, each component has been normalized by a reference measurement without sample. This ensures that the  amplitude vector of the incident wave is represented by $\textbf{\emph{E}}_0=(1,1)$.

The results are presented in Fig.\ \ref{fig:results}, where we show the magnitude of the
amplitude transmittance for each component $|E_x|$, $|E_y|$ (Fig.\ \ref{fig:results}(a)),
the phase delay $\Delta\phi$ between the $E_x$ and $E_y$ component (Fig.\
\ref{fig:results}(b)), the overall intensity transmittance $S=|\textbf{\emph{E}}|^2/2$ and the
flattening $f$ (Fig.\ \ref{fig:results}(c)) in dependence of the frequency for a QWP (left column) and a HWP (right column).

Table \ref{tab:table1} summarizes the quantitative results for the center frequency $\nu_0$, the flattening $f$, the intensity transmittance $S$, the phase delay $\Delta\phi$ , the spectral bandwidth $\Delta \nu$ and the variation of the intensity ratio $\Delta S$ within the spectral bandwidth as observed for the QWP and HWP. Thereby, the spectral bandwidth for the wave plates, as indicated by the shadowed regions in Fig.\ \ref{fig:results}, was defined as the frequency range around the center frequency where the flattening deviated less than 20\% from the optimum value (QWP: $f=0$, QWP: $f=1$). Due to the inherently existent strong dispersion in doubly-resonant negative index metamaterials, the phase difference between the two orthogonal polarization components of the wave changed rapidly (see Fig.\ \ref{fig:results}). For this reason, only narrow-band wave plates can be realized by the proposed method. On the other hand, we achieved a high transmission for parallel and perpendicular wave polarization over a wide frequency band.
\begin{table}[t]
\caption{\label{tab:table1} Center frequency $\nu_0$, flattening $f$, intensity transmittance $S$, phase delay $\Delta\phi$, spectral bandwidth $\Delta f$  and variation of the intensity transmittance $\Delta S$ for QWP and HWP}
\begin{ruledtabular}
\begin{tabular}{c|cccccc}
& $\nu_0 [THz]$ & $f$ & $S$ & $\Delta\phi$ & $\Delta \nu [GHz]$ & $\Delta S$ \\
\hline
QWP & 1.3 & 0.09 & 0.74 & 88.9 & 38 & 0.68 - 0.75\\
HWP & 1.34 & 1.00 & 0.58 & 180.6 & 38 & 0.56 - 0.58 \\
\end{tabular}
\end{ruledtabular}
\end{table}
\begin{figure}[]
	\begin{center}
		\includegraphics[width=\columnwidth]{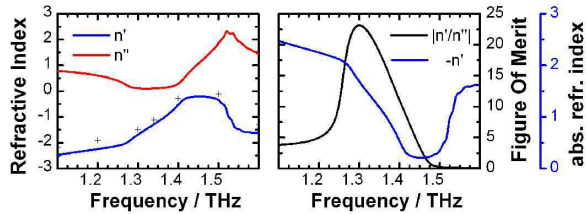}
    \end{center}
	\caption{Real part $n'$, imaginary part $n''$ and figure of merit $|n'/n''|$ retrieved from the transmission and reflection data. The crosses indicate $n'$ calculated from the phase advance in the metamaterial.}
	\label{fig:index}
\end{figure}

To get a deeper insight in the functionality of the metamaterial wave plates, we retrieved the effective material parameters from the simulated reflection and transmission data \cite{chen:2004}. We additionally calculated the real part of the refractive index from the phase advance in the material (indicated by crosses in Fig.\ \ref{fig:index}) to eliminate branch ambiguities in the retrieval procedure. For the numerical calculations we used the time domain solver of CST Microwave studio and modeled BCB as a dielectric with  a permittivity of $\varepsilon_\textrm{BCB}=2.75$ and a loss tangent of $\tan(\delta)=0.008$. We observed that the retrieved data were independent of the number of unit cells in the propagation direction and thus identical for the HWP and the QWP. The resulting refractive index and the figure of merit $ FOM=|n'/n''|$ are plotted in Fig.\ \ref{fig:index}. It is remarkable, that the highest observed figure of merit was $FOM=23$ for a refractive index of $n'=-1.7$ at 1.3 THz. For the HWP, we retrieved a phase difference between the two orthogonally polarized waves of $\Delta \phi=560{^\circ}$ from the refractive indices $n'_{||}=-1.33$ and $n'_\bot = 1.85$ at a center frequency of 1.34~THz. This value can only be interpreted as a rough estimate due to the high sensitivity of $\Delta \phi$ to the relative values of the refractive indices and the absolute thickness of the sample. Nevertheless, it evidences that, despite of the subwavelength metamaterial thickness of only 110~$\mu$m, a very large phase shift of approximately 3 times $180{^\circ}$ could be achieved originating from the strong birefringence in the medium. The same holds for the quarter wave plate, where we estimated a phase shift of 3 times $90{^\circ}$ for a 55 $\mu$m thin plate at a frequency of 1.3~THz. This is in agreement with the modulo $360{^\circ}$ phase delay data obtained from the experimental measurements.

In conclusion, we have experimentally and theoretically demonstrated that strongly birefringent metamaterials which provide refractive indices of opposite sign for two orthogonal field polarizations can serve as efficient, subwavelength thin THz wave plates. The described quarter- and half wave plates based on a (multilayer) wire pair design offered an intensity transmittance higher than 74\% and 58\%, respectively. As a drawback, the wave plates only operated accurately in a relatively narrow frequency band in the order of several 10 GHz. Furthermore, the figure of merit of the metamaterial approached a peak value of $FOM=23$ at 1.3 THz which corresponds to one of the highest reported at THz frequencies. The presented results evidence that negative index materials enter an application stage in terms of optical components for the THz technology.

We would like to acknowledge the Nano+Bio Center in Kaiserslautern, Germany for support and valuable discussions.


\end{document}